\pdfoutput=1
\documentclass[sigconf]{acmart}

\setcopyright{none}
\renewcommand\footnotetextcopyrightpermission[1]{}

\usepackage[utf8]{inputenc} 
\usepackage[T1]{fontenc}    
\usepackage{hyperref}       
\usepackage{url}            
\usepackage{booktabs}       
\usepackage{nicefrac}       
\usepackage{microtype}      
\usepackage{xcolor}         
\usepackage{tikz}
\usepackage{enumitem}
\usepackage[normalem]{ulem}
\usepackage{graphicx}
\usepackage{multirow}
\usepackage{xspace}
\usepackage{makecell}
\usepackage{tablefootnote}
\usepackage{soul}
\usepackage{xspace}
\usepackage{makecell}
\usepackage{balance}

\newcommand{\insertmiraclresults}[0]{
\begin{table*}[]
\resizebox{0.6\textwidth}{!}{
\begin{tabular}{lccccccc}
\toprule
\multirow{2}{*}{\textbf{Language}} & \multirow{2}{*}{\textbf{ISO}}  & \multicolumn{3}{c}{\textbf{nDCG@10}} & \multicolumn{3}{c}{\textbf{Recall@100}} \\
& & \textbf{BM25} & \textbf{mDPR} & \textbf{Hybrid} & \textbf{BM25} & \textbf{mDPR} & \textbf{Hybrid} \\
\cmidrule(lr){1-2} \cmidrule(lr){3-5} \cmidrule(lr){6-8} 
Arabic & \Ar & 0.481 & 0.499 & 0.673 & 0.889 & 0.841 & 0.941 \\
Bengali & \Bn & 0.508 & 0.443 & 0.654 & 0.909 & 0.819 & 0.932 \\
English & \En & 0.351 & 0.394 & 0.549 & 0.819 & 0.768 & 0.882 \\
Spanish & \Es & 0.319 & 0.478 & 0.641 & 0.702 & 0.864 & 0.948 \\
Persian & \Fa & 0.333 & 0.480 & 0.594 & 0.731 & 0.898 & 0.937 \\
Finnish & \Fi & 0.551 & 0.472 & 0.672 & 0.891 & 0.788 & 0.895 \\
French & \Fr & 0.183 & 0.435 & 0.523 & 0.653 & 0.915 & 0.965 \\
Hindi & \Hi & 0.458 & 0.383 & 0.616 & 0.868 & 0.776 & 0.912 \\
Indonesian & \Id & 0.449 & 0.272 & 0.443 & 0.904 & 0.573 & 0.768 \\
Japanese & \Ja & 0.369 & 0.439 & 0.576 & 0.805 & 0.825 & 0.904 \\
Korean & \Ko & 0.419 & 0.419 & 0.609 & 0.783 & 0.737 & 0.900 \\
Russian & \Ru & 0.334 & 0.407 & 0.532 & 0.661 & 0.797 & 0.874 \\
Swahili & \Sw & 0.383 & 0.299 & 0.446 & 0.701 & 0.616 & 0.725 \\
Telugu & \Te & 0.494 & 0.356 & 0.602 & 0.831 & 0.762 & 0.857 \\
Thai & \Th & 0.484 & 0.358 & 0.599 & 0.887 & 0.678 & 0.823 \\
Chinese & \Zh & 0.180 & 0.512 & 0.526 & 0.560 & 0.944 & 0.959 \\
\cmidrule(lr){1-2} \cmidrule(lr){3-5} \cmidrule(lr){6-8} 
Average & & 0.393 & 0.415 & 0.578 & 0.787 & 0.788 & 0.889 \\
\bottomrule
\end{tabular}
}
\vspace{0.2cm}
\caption{Baseline results on the development set of \miracl:\ BM25 uses the implementation in Anserini with default parameters and language-specific analyzers;
mDPR is fine-tuned on the MS MARCO Passage dataset and applied to each language in a zero-shot manner;
Hybrid combines BM25 and mDPR scores.
}
\label{tab:baselines}
\end{table*}
}

\newcommand{\insertmiraclstats}[0]{
    \begin{table*}[]
    \resizebox{1.0\textwidth}{!}{
    \begin{tabular}{lcrrrrrrrrrrc}
    \toprule
    \multirow{2}{*}{\textbf{Language}} & \multirow{2}{*}{\textbf{ISO}} & \multicolumn{2}{c}{\textbf{Train}} & \multicolumn{2}{c}{\textbf{Dev}} & \multicolumn{2}{c}{\textbf{Test-A}} & \multicolumn{2}{c}{\textbf{Test-B}} & \multicolumn{1}{l}{\multirow{2}{*}{\bf \# Passages}} & \multicolumn{1}{l}{\multirow{2}{*}{\bf \# Articles}} & \multirow{2}{*}{\textbf{\makecell{In \\ \mrtydi?}}} \\
    & & \multicolumn{1}{r}{\textbf{\# Q}} & \multicolumn{1}{r}{\textbf{\# J}} & \multicolumn{1}{r}{\textbf{\# Q}} & \multicolumn{1}{r}{\textbf{\# J}} & \multicolumn{1}{r}{\textbf{\# Q}} & \multicolumn{1}{r}{\textbf{\# J}} & \multicolumn{1}{r}{\textbf{\# Q}} & \multicolumn{1}{r}{\textbf{\# J}} & \multicolumn{1}{r}{} & \multicolumn{1}{r}{} \\
     \cmidrule(lr){1-2} \cmidrule(lr){3-4} \cmidrule(lr){5-6} \cmidrule(lr){7-8} \cmidrule(lr){9-10} \cmidrule(lr){11-12} \cmidrule(lr){13-13}
Arabic & {\Ar} & 3,495 & 25,382 & 2,896 & 29,197 & 936 & 9,325 & 1,405 & 14,036 & 2,061,414 & 656,982 & \tick \\
Bengali & {\Bn} & 1,631 & 16,754 & 411 & 4,206 & 102 & 1,037 & 1,130 & 11,286 & 297,265 & 63,762 & \tick \\
English & {\En} & 2,863 & 29,416 & 799 & 8,350 & 734 & 5,617 & 1,790 & 18,241 & 32,893,221 & 5,758,285 & \tick \\
Spanish & {\Es} & 2,162 & 21,531 & 648 & 6,443 & -- & -- & 1,515 & 15,074 & 10,373,953 & 1,669,181 & \cross \\
Persian & {\Fa} & 2,107 & 21,844 & 632 & 6,571 & -- & -- & 1,476 & 15,313 & 2,207,172 & 857,827 & \cross \\
Finnish & {\Fi} & 2,897 & 20,350 & 1,271 & 12,008 & 1,060 & 10,586 & 711 & 7,100 & 1,883,509 & 447,815 & \tick \\
French & {\Fr} & 1,143 & 11,426 & 343 & 3,429 & -- & -- & 801 & 8,008 & 14,636,953 & 2,325,608 & \cross\\
Hindi & {\Hi} & 1,169 & 11,668 & 350 & 3,494 & -- & -- & 819 & 8,169 & 506,264 & 148,107 & \cross \\
Indonesian & {\Id} & 4,071 & 41,358 & 960 & 9,668 & 731 & 7,430 & 611 & 6,098 & 1,446,315 & 446,330 & \tick \\
Japanese & {\Ja} & 3,477 & 34,387 & 860 & 8,354 & 650 & 6,922 & 1,141 & 11,410 & 6,953,614 & 1,133,444 & \tick \\
Korean & {\Ko} & 868 & 12,767 & 213 & 3,057 & 263 & 3,855 & 1,417 & 14,161 & 1,486,752 & 437,373 & \tick \\
Russian & {\Ru} & 4,683 & 33,921 & 1,252 & 13,100 & 911 & 8,777 & 718 & 7,174 & 9,543,918 & 1,476,045 & \tick \\
Swahili & {\Sw} & 1,901 & 9,359 & 482 & 5,092 & 638 & 6,615 & 465 & 4,620 & 131,924 & 47,793 & \tick \\
Telugu & {\Te} & 3,452 & 18,608 & 828 & 1,606 & 594 & 5,948 & 793 & 7,920 & 518,079 & 66,353 & \tick \\
Thai & {\Th} & 2,972 & 21,293 & 733 & 7,573 & 992 & 10,432 & 650 & 6,493 & 542,166 & 128,179 & \tick \\
Chinese & {\Zh} & 1,312 & 13,113 & 393 & 3,928 & -- & -- & 920 & 9,196 & 4,934,368 & 1,246,389 & \cross \\
     \cmidrule(lr){1-2} \cmidrule(lr){3-4} \cmidrule(lr){5-6} \cmidrule(lr){7-8} \cmidrule(lr){9-10} \cmidrule(lr){11-12} \cmidrule(lr){13-13}
Total & & 40,203 & 343,177 & 13,071 & 126,076 & 7,611 & 76,544 & 16,362 & 164,299 & 90,416,887 & 16,909,473 \\
    \midrule
    \multicolumn{2}{l}{Surprise Language 1} & ? & ? & ?& ?& ?& ?& ?& ?& ? & ? & \cross \\
    \multicolumn{2}{l}{Surprise Language 2} & ? & ? & ?& ?& ?& ?& ?& ?& ? & ? & \cross\\
    \bottomrule
    \end{tabular}
    }
   \vspace{0.2cm}
   \caption{
        Descriptive statistics for each language, split combination in \miracl:\ \#~Q denotes the number of queries; \#~J denotes the number of judgments (both relevant and non-relevant).
        Statistics of each Wikipedia corpus are also provided:\ \#~Passages denotes the number of passages in each language;
        \#~Articles denotes the number of Wikipedia articles from which the passages were drawn.
        The final column indicates if the language is contained in \mrtydi.
        \miracl encompasses 18 languages in total:\ 16 of which are known, with 2 ``surprise'' languages whose identities will be revealed in the future.
    }
    \label{tab:miracl-stats} 
    \end{table*}
}

\newcommand\Ar{\texttt{ar}\xspace}
\newcommand\Bn{\texttt{bn}\xspace}
\newcommand\En{\texttt{en}\xspace}
\newcommand\Es{\texttt{es}\xspace}
\newcommand\Fa{\texttt{fa}\xspace}
\newcommand\Fi{\texttt{fi}\xspace}
\newcommand\Fr{\texttt{fr}\xspace}
\newcommand\Hi{\texttt{hi}\xspace}
\newcommand\Id{\texttt{id}\xspace}
\newcommand\Ja{\texttt{ja}\xspace}
\newcommand\Ko{\texttt{ko}\xspace}
\newcommand\Ru{\texttt{ru}\xspace}
\newcommand\Sw{\texttt{sw}\xspace}
\newcommand\Te{\texttt{te}\xspace}
\newcommand\Th{\texttt{th}\xspace}
\newcommand\Zh{\texttt{zh}\xspace}

\newcommand{\ignore}[1]{}

\newcommand{\cross}[0]{\textcolor{black}{$\times$}}
\newcommand{\tick}[0]{\textcolor{black}{\checkmark}}

\newcommand{\miracl}[0]{MIRACL\xspace}
\newcommand\tydi{\textsc{TyDi QA}\xspace}
\newcommand\mrtydi{Mr.~\textsc{TyDi}\xspace}

\begin{document}

\title{Making a \miracl: Multilingual Information Retrieval \\ Across a Continuum of Languages}

\author{
Xinyu Zhang$^{*1}$, Nandan Thakur$^{*1}$, 
Odunayo Ogundepo$^1$, 
Ehsan Kamalloo$^{2,3}$, \\
David Alfonso-Hermelo$^3$,
Xiaoguang Li$^3$,
Qun Liu$^3$,
Mehdi Rezagholizadeh$^3$,
Jimmy Lin$^1$
}
\thanks{$^*$ Equal Contribution}

\affiliation{\vspace{0.1cm}
$^1$ David R. Cheriton School of Computer Science, 
University of Waterloo \country{Canada}  \\
$^2$  Department of Computing Science, University of Alberta \country{Canada} \qquad $^3$ Huawei Noah's Ark Lab
}

\renewcommand{\shortauthors}{}
\pagestyle{empty}

\begin{abstract}
MIRACL (Multilingual Information Retrieval Across a Continuum of Languages) is a multilingual dataset we have built for the WSDM 2023 Cup challenge that focuses on {\it ad hoc} retrieval across 18 different languages, which collectively encompass over three billion native speakers around the world.
These languages have diverse typologies, originate from many different language families, and are associated with varying amounts of available resources---including what researchers typically characterize as high-resource as well as low-resource languages.
Our dataset is designed to support the creation and evaluation of models for monolingual retrieval, where the queries and the corpora are in the same language. 
In total, we have gathered over 700k high-quality relevance judgments for around 77k queries over Wikipedia in these 18 languages, where all assessments have been performed by native speakers hired by our team.
Our goal is to spur research that will improve retrieval across a continuum of languages, thus enhancing information access capabilities for diverse populations around the world, particularly those that have been traditionally underserved.
This overview paper describes the dataset and baselines that we share with the community.
The MIRACL website is live at \url{http://miracl.ai/}.

\end{abstract}

\maketitle

\begin{figure}[t]
\includegraphics[trim=0 5 0 0,clip,width=\columnwidth]{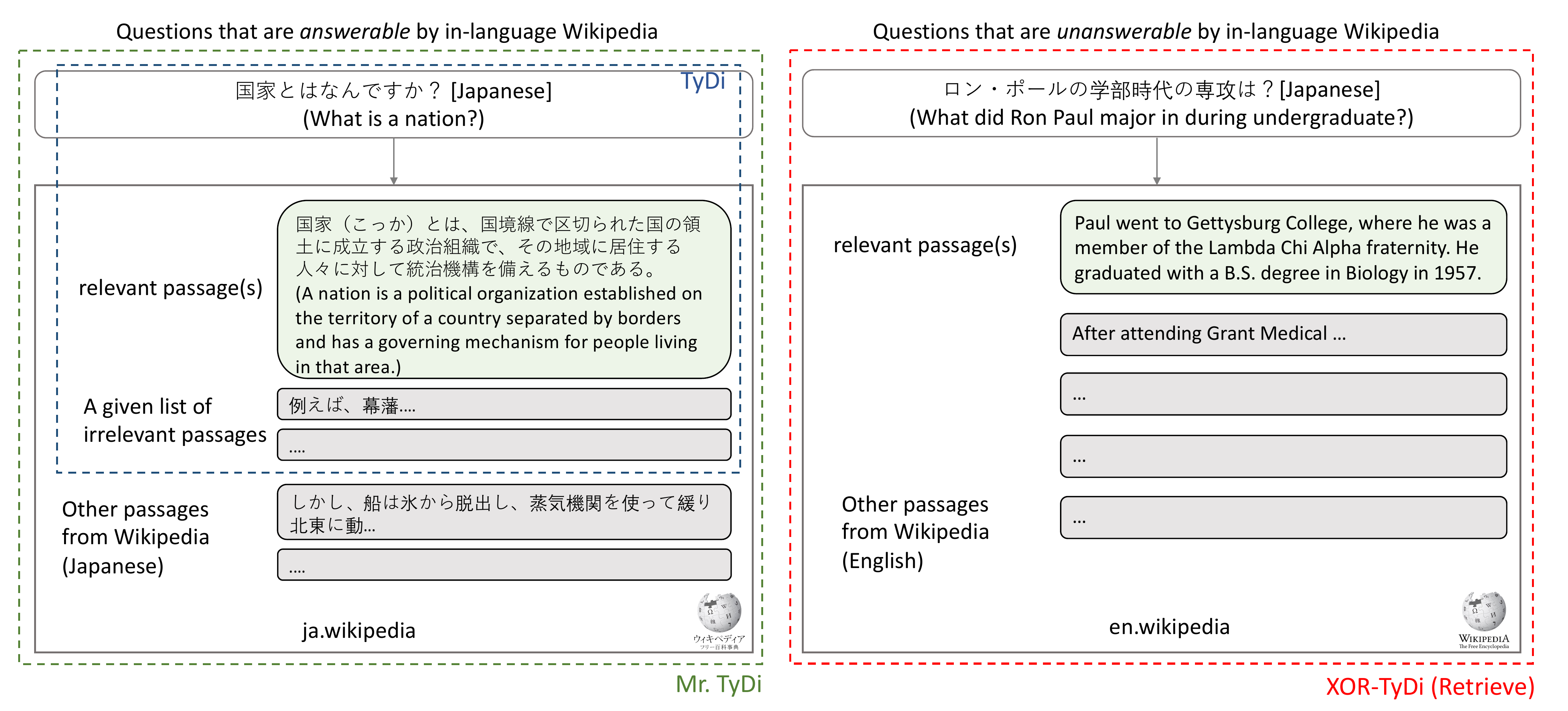}
\caption{
Examples of annotated query--passage pairs in Thai (\Th) from \miracl. Queries are generated by native speakers and passages are prepared from the corresponding Wikipedia in the same language (e.g., \url{th.wikipedia.org} in this example).}
\label{fig:miracl-example}
\end{figure}

\section{Introduction}

Information access is a fundamental human right.
Specifically, the Universal Declaration of Human Rights by the United Nations articulates that ``everyone has the right to freedom of opinion and expression'', which includes the right ``to seek, receive and impart information and ideas through any media and regardless of frontiers'' (Article 19).
Information access capabilities such as search, question answering, and recommendation are important technologies for safeguarding these ideals. 

With the advent and dominance of deep learning and approaches based on neural networks (particularly transformer-based large language models) in natural language processing, information retrieval, and beyond, the importance of large datasets as drivers of progress is well understood~\cite{Lin_etal_2021_ptr4tr}.
For retrieval models in English, the MS MARCO datasets~\cite{msmarco,Craswell_etal_SIGIR2021_perspectives,Lin_etal_SIGIR2022} have had a transformative impact in advancing the field.
Similarly, for question answering (including the so-called ``open-domain'' retrieval-based variant), there exist many resources in English, such as SQuAD~\citep{rajpurkar-etal-2016-squad}, TriviaQA~\citep{joshi-etal-2017-triviaqa}, and Natural Questions~\citep{nq}.
We have recently witnessed efforts in building resources for non-English languages, for example,  CLIRMatrix~\cite{sun-duh-2020-clirmatrix}, XTREME~\citep{xtreme}, MKQA \cite{longpre-etal-2021-mkqa}, mMARCO~\citep{bonifacio2021mmarco}, {\tydi}~\citep{clark-etal-2020-tydi}, XOR-\textsc{TyDi}~\cite{asai-etal-2021-xor}, and \mrtydi~\citep{mrtydi}.
These initiatives complement cross-lingual retrieval evaluations from TREC, CLEF, NTCIR, and FIRE that date back many years, largely focused on specific language pairs.
Nevertheless, there remains a paucity of resources for languages beyond English.
Existing datasets are far from sufficient to fully develop information access capabilities for the 7000+ languages spoken on our planet~\cite{joshi-etal-2020-state}. 
Our goal is to take a small step towards addressing these issues.

To stimulate further advances in multilingual retrieval, we have built the \miracl dataset, comprising human-annotated passage-level relevance judgments on Wikipedia for 18 languages, totaling over 700k query--passage pairs on 77k queries.
These languages---Arabic (\Ar), Bengali (\Bn), English (\En), Spanish (\Es), Farsi (\Fa), Finnish (\Fi), French (\Fr), Hindi (\Hi), Indonesian (\Id), Japanese (\Ja), Korean (\Ko), Russian (\Ru), Swahili (\Sw), Telugu (\Te), Thai (\Th), Chinese (\Zh), and two ``surprise'' languages to be revealed later---are written using 11 distinct scripts, originate from 10 different language families, and collectively encompass over three billion native speakers around the world.
They include what the research community would typically characterize as high-resource languages as well as low-resource languages.

\insertmiraclstats{}

Figure~\ref{fig:miracl-example} shows an example of a query from \miracl in Thai along with a relevant and a non-relevant passage.
Along with the \miracl dataset, our broader efforts (i.e., the ``\miracl project'') include organizing a WSDM 2023 Cup challenge\footnote{\url{https://www.wsdm-conference.org/2023/program/wsdm-cup}} that provides a common evaluation methodology, a leaderboard, and a venue for a competition-style event with prizes.
To provide starting points that the community can rapidly build on, we also share reproducible BM25, mDPR, and hybrid baselines as part of our Pyserini~\cite{Lin_etal_SIGIR2021_Pyserini} and Anserini~\cite{Yang_etal_JDIQ2018} toolkits.

This paper focuses on providing a descriptive overview of the \miracl dataset to coincide with our initial data release.
It is our intention to periodically update this document with additional details about the WSDM 2023 Cup challenge as well as the broader \miracl project over time.

\section{Dataset Overview}
\label{sec:dataset}

\miracl (Multilingual Information Retrieval Across a Continuum of Languages) is a multilingual retrieval dataset that spans 18 different languages (see Table \ref{tab:miracl-stats}).
More precisely, the task we model is the standard {\it ad hoc} retrieval task as defined by the information retrieval community, where given a corpus $\mathcal{C}$, the system's task is to return for a given $q$ an ordered list of top-$k$ documents from $\mathcal{C}$ that maximizes some standard metric of quality such as nDCG.
In our formulation, a query $q$ is a well-formed natural language question in some language $\mathcal{L}_n$ (one of 18) and the documents draw from a snapshot of Wikipedia in the same language $\mathcal{C}_n$ that has been pre-segmented into passages (and thus each passage has a fixed unique identifier).

Thus, our focus is monolingual retrieval across diverse languages, where the queries and the corpora are in the same language (e.g., Thai queries searching Thai documents), as opposed to {\it cross-lingual} retrieval, where the queries and the corpora are in {\it different} languages (e.g., searching a Swahili corpus with Arabic queries).
As a terminological note, consistent with the parlance in information retrieval, we use the term ``document'' to refer generically to the unit of retrieval, even though in this case the ``documents'' are in actuality passages from Wikipedia.

In total, we have gathered over 700k manual relevance judgments (i.e., query--passage pairs) for around 77k queries across Wikipedia in these 18 languages, where all assessments have been performed by native speakers.
Section~\ref{sec:construction} describes the annotation process in more detail.
For each language, these relevance judgments are divided into a training set, a development set, and two test sets (that we call test-A and test-B); more details below.
The \miracl dataset is released under an Apache 2.0 License.
To evaluate the quality of the system output, we use standard information retrieval metrics such as nDCG at a fixed cutoff and recall at a fixed cutoff.

The \miracl dataset is built on the \mrtydi multilingual retrieval dataset~\cite{mrtydi}, which is in turn built on the \tydi dataset~\cite{clark-etal-2020-tydi}.
Beyond the 11 languages originally covered by \mrtydi and \tydi, we included 7 additional languages.
Details below:

\begin{itemize}[leftmargin=*]

\item 
\textbf{Existing (Known) Languages:} 
\mrtydi and \tydi cover 11 languages:\ Arabic (\Ar), Bengali (\Bn), English (\En), Finnish (\Fi), Indonesian (\Id), Japanese (\Ja), Korean (\Ko), Russian (\Ru), Swahili (\Sw), Telugu (\Te), and Thai (\Th).
We take advantage of existing queries in these languages as a starting point and provide ``denser'' annotations with respect to Wikipedia passages.
That is, each query in \mrtydi has on average only a single positive (relevant) passage~\cite{mrtydi}.
For existing \mrtydi queries, \miracl provides richer judgments (i.e., more positive as well as negative labels).
Beyond more comprehensive annotations of existing \mrtydi queries, \miracl also includes new queries (i.e., created from scratch) that have been annotated with the same methodology.

\smallskip \noindent 
\item
\textbf{New (Known) Languages:}
To the languages in \mrtydi and \tydi, we add 5 new (known) languages:\ Hindi (\Hi), Spanish (\Es), French (\Fr), Farsi (\Fa), and Chinese (\Zh).
For these languages, there are no existing resources to build on, and thus all data are generated from scratch.

\smallskip \noindent 
\item
\textbf{New (Surprise) Languages:}
Finally, there are two new (surprise) languages whose identities are presently hidden, but will be revealed at a later date as part of the WSDM 2023 Cup challenge.
The construction of data for the surprise languages follows the same procedure as the new (known) languages.

\end{itemize}

\smallskip
\noindent
At a high-level, \miracl contains two classes of queries:\ those inherited from \mrtydi and those that were created from scratch.
For the first class of queries, the splits in \miracl align with the splits from \mrtydi.
In more detail:

\begin{itemize}[leftmargin=*]

\item {\bf Training sets:}
For the 11 existing \mrtydi languages, the training sets comprise subsets of the queries from the \mrtydi training sets (originating from \tydi).
The main difference is that \miracl provides richer annotations for more passages.
However, our annotators were not able to find relevant passages for some queries from \mrtydi; we call these ``invalid'' queries, and they were removed from the \miracl training sets (see Section~\ref{sec:construction:discussion}).
For this reason, there are fewer queries in the \miracl training sets than in their corresponding \mrtydi splits.
For the new languages (both known and surprise), the training data comprise entirely of newly generated queries.

\smallskip \noindent 
\item {\bf Development sets:}
Similar to the training sets, the \miracl development sets align with  \mrtydi for existing languages, but with the ``invalid'' queries discarded (see above).
For the new languages, the development sets comprise entirely of newly generated queries.

\smallskip \noindent 
\item {\bf Test-A sets:}
The test-A sets exist {\it only} for the existing \mrtydi languages and align with the test sets in \mrtydi (once again, with ``invalid'' queries discarded).

\smallskip \noindent 
\item {\bf Test-B sets:}
For all languages, the test-B sets are comprised entirely of new queries that have never been released before (compared to test-A sets, which ultimately draw from \tydi and thus have been publicly available for quite some time now).
These queries will be used in the final evaluation for the WSDM 2023 Cup challenge, and can be viewed as a {\it true} held-out test set.

\end{itemize}

\noindent Detailed statistics for \miracl are shown in Table \ref{tab:miracl-stats}.
For each language, split combination described above, we report the number of queries (\# Q) and the number of relevance judgments (\# J), which includes both relevant as well as non-relevant passages.
The size of each Wikipedia corpus is reported in terms of the number of passages (\# Passages) and the number of articles (\# Articles) from which they were derived.
As the relevance judgements were provided on query--passage pairs, the total number of articles is provided primarily as a reference.
The final column indicates whether the language is part of \mrtydi and \tydi, or is one of the new languages.

It is our expectation that there are sufficient examples in the training and development sets of \miracl to train (specifically, fine-tune) transformer-based retrieval models.
We believe that \miracl will provide a valuable dataset for building and evaluating dense retrieval models such as DPR~\citep{karpukhin-etal-2020-dense}, late-interaction models such as ColBERT~\citep{Khattab_Zaharia_SIGIR2020}, as well as reranking models such as monoBERT~\cite{monobert} and monoT5~\cite{monot5}.
For these classes of retrieval models, there is an emerging thread of research~\cite{shi-etal-2020-cross,MacAvaney_etal_ECIR2020,Nair_etal_ECIR2022,mrtydi-ood} we hope that \miracl will further catalyze.

\section{Dataset Construction} 
\label{sec:construction}

To build \miracl, we hired native speakers of each language as annotators to provide high-quality queries and judgments.
At a high level, our workflow comprised two phases:\
First, the annotators were asked to generate well-formed natural language questions based on ``prompt'' paragraphs.
Then, they were asked to assess the relevance of the top-$k$ query---passage pairs produced by an ensemble baseline retrieval system.

An important feature of \miracl is that our dataset was {\it not} constructed via crowd-sourced workers, unlike other previous efforts such as SQuAD~\cite{rajpurkar-etal-2016-squad}.
Instead, we hired 31 annotators (both part-time and full-time) across all languages.
Each annotator was interviewed prior to being hired and was verified to be a native speaker of the language they were working in.
Our team created a consistent onboarding process that included training the annotators on exactly the tasks they were asked to perform.
Throughout the annotation process, we checked randomly sampled data to monitor annotation quality.
We believe that this design decision yielded a higher quality dataset than could have been obtained by crowd-sourcing means.
Our team began interviewing annotators in mid-April 2022; dataset construction began in earnest in late April 2022 and continued until the end of September 2022.

\subsection{Corpora Preparation}

For each \miracl language, we prepared a pre-segmented passage corpus from a raw Wikipedia dump.
For the existing languages in \mrtydi, we used exactly the same raw Wikipedia dump as \mrtydi and \tydi (from January 1, 2019 for Thai and from February 1, 2019 for the others).
For the new languages, we used the versions released on March 1, 2022.
We parsed the Wikipedia articles using WikiExtractor\footnote{\url{https://github.com/attardi/wikiextractor}}
and segmented them into passages based on natural discourse units using two consecutive newlines in the wiki markup as the delimiter.

Each passage is given a unique identifier based on an \texttt{X\#Y} schema, where \texttt{X} refers to a unique Wikipedia article and \texttt{Y} refers to the passage (numbered sequentially) within that article.
Thus, it is possible for a system to reconstruct the passage/article relationships.
To be clear, however, our task is to retrieve relevant passages, and thus query--passage pairs form the basic annotation unit.
The corpus for each language is distributed in JSON lines format.

A key difference between \miracl and \mrtydi is the manner in which the corpus was prepared.
Since \mrtydi derived passage-level relevance annotations from \tydi, it retained exactly those same passages.
However, since \tydi was not originally designed for retrieval, it did not provide consistent passage segmentation for all of Wikipedia.
Thus, the \mrtydi corpora comprised a mix of \tydi passages and custom segments that were heuristically adjusted to ``cover'' the entire raw Wikipedia dump.
This inconsistent segmentation is a weakness of \mrtydi that we rectified in the design of \miracl.
The downside, unfortunately, is that annotated passages from \mrtydi may no longer exist in the \miracl corpora.
Thus, to take advantage of existing annotations, we ``projected'' relevant passages from \mrtydi onto our newly prepared corpora.
This process is described below.

\subsection{Annotation Workflow}
\label{sec:construction:workflow}

\miracl was created using a two-phase process adapted from  {\tydi}, which was in turn built on best practices derived from previous work; see discussion by \citet{clark-etal-2020-tydi}.
The two phases are \textit{query generation} and \textit{relevance assessment}, detailed below:

\smallskip
\noindent {\bf Query Generation.}
In the first phase, annotators were shown ``prompts'' from Wikipedia that provide contexts to elicit queries. 
The prompts were extracted from the first 100 words of randomly selected Wikipedia articles.
To generate high-quality queries, we asked that the annotators avoid generating queries that are answerable by the prompts themselves.
Annotators were asked to generate well-formed natural language questions that are likely (in their opinion) to have precise, unambiguous answers.
They could skip prompts that they did not find to be ``inspiring''.

Note that in this phase, annotators were asked to generate questions in batch based on the prompts.
As a result of this workflow, at this point, the annotators have not yet examined any retrieval results (i.e., passages from Wikipedia), and it could be the case that their questions can not be readily answered by information contained in the corpus.
We discuss this case in Section~\ref{sec:construction:discussion}.

\smallskip
\noindent {\bf Relevance Assessment.}
In the second phase, for each query from the previous phase, we asked the annotators to judge the binary relevance of the top-$k$ candidate passages ($k=10$) from a baseline ensemble retrieval system that combined three separate models:

\begin{itemize}[leftmargin=*]

\item Lexical matching with BM25:\  
We used the Anserini~\cite{Yang_etal_JDIQ2018} implementation based on the open-source Lucene search library with default parameters and the corresponding language-specific analyzer provided by Lucene. 

\item A single-representation bi-encoder model, mDPR~\citep{karpukhin-etal-2020-dense, mrtydi}: 
We trained an mDPR model using the Tevatron toolkit~\cite{tevatron}, where the model was initialized from an mBERT checkpoint\footnote{\texttt{bert-case-multilingual-base} on HuggingFace.}
and then fine-tuned using the training set of MS MARCO Passage~\cite{mrtydi-ood}.
Retrieval was performed in a zero-shot manner.

\item A late-interaction model, mColBERT~\citep{Khattab_Zaharia_SIGIR2020,bonifacio2021mmarco}:
We trained mColBERT using the authors' official repository.\footnote{\url{https://github.com/stanford-futuredata/ColBERT\#colbertv1}}
Similar to mDPR, the model was initialized from the same mBERT checkpoint as mDPR and then fine-tuned on MS MARCO Passage.
As with mDPR, retrieval was performed in a zero-shot manner.

\end{itemize}

\noindent For each query, we retrieved the top-$k$ passages ($k=1000$) using each model.
We then performed ensemble fusion by first normalizing all retrieval scores to the range $[0, 1]$ and then averaging the scores.
A final ranked list was then generated from these new scores.
Based on initial experiments, we found that annotating the top 10 passages for each query yielded a good balance in terms of obtaining diverse passages and efficiently utilizing annotator effort.

For queries from the 11 existing languages in \mrtydi, we augmented the set of passages to be assessed with ``projected'' relevant passages from \mrtydi.
This allowed us to take advantage of existing annotations transparently in our workflow.
To accomplish this, we used relevant passages from \mrtydi as queries to search the corresponding \miracl corpus using BM25.
As the passages in \miracl and \mrtydi differ in terms of segmentation but not content, the top retrieved passages are likely to have substantial overlap with the annotated passage in \mrtydi (and hence are also likely to be relevant).\footnote{We cannot simply assume that highly similar passages from \miracl are also relevant.}

We used a simple heuristic to determine how many of these results to re-assess.
If the score of the top retrieved passage from \miracl is 50\% higher than the score of the passage ranked second, we are more confident that the top passage is a good match for the original relevant passage.
In this case, we only add the top passage to the set of candidates that the assessor considers.
Otherwise, we add the top 5 passages.
Note that these ``projected'' passages are not specially identified from the perspective of the annotator.
In all cases, they simply receive a set of passages to judge per query, without any explicit knowledge where the passages came from.

\subsection{Fold Creation and Data Release}

During the annotation process, all queries followed the same workflow.
After the annotation process concluded, we divided \miracl into training sets, development sets, test-A sets, and test-B sets, as described in Section~\ref{sec:dataset}.

For existing languages, the training, development, and test-A sets align with the training, development, and test sets in \mrtydi,
and the test-B sets are formed by the newly generated questions.
For the new (known) languages, all generated queries were split into training, development, and test-B sets with a split ratio of $50\%$, $15\%$, and $35\%$.
Note that there are no test-A sets for these languages.
The identity of the surprise languages and associated relevance judgments are currently kept secret and will be released as part of the WSDM 2023 Cup challenge.

Presently, all training and development sets have been released to the public, including both queries and relevance judgments.
There are no current plans to release the relevance judgments for the test-A and test-B sets.

\insertmiraclresults

\subsection{Discussion}
\label{sec:construction:discussion}

Since \miracl inherits from \mrtydi, which was in turn built on \tydi, it is worthwhile to discuss important differences in the annotation workflow.

At the core, {\tydi}~\citep{clark-etal-2020-tydi} is {\it not} a dataset to evaluate retrieval.
Rather, it is best characterized as a dataset for the so-called machine reading comprehension task, where the system is given (query, text) pairs and asked to identify the answer within the text.
In \tydi, candidate passages for annotation are selected from only the top Wikipedia article based on a Google search.
In contrast, in \miracl we draw candidate passages from across all Wikipedia articles.
This makes relevant passages in \miracl more diverse.

Since we asked annotators to assess the top-$k$ candidate passages per query, \miracl annotations are richer and more diverse than \mrtydi judgments (which were mechanistically generated from \tydi as no new annotations were performed).
Furthermore, we believe that explicitly judged negative examples are quite valuable (compared to, for example, implicit negatives in MS MARCO sampled from BM25 results), as recent work has demonstrated the importance of so-called ``hard negative'' mining~\cite{xiong-etal-2021-ance-iclr}.
Since the candidate passages already come from an ensemble comprising baseline neural models, these are particularly suitable examples for various contrastive learning techniques.

One important detail worth discussing:\
Since the query generation and relevance assessment phases were decoupled, for some fraction of queries in each language, annotators were not able to identify any relevant passage in the pairs provided to them.
For \miracl, we aimed to have at least one relevant passage per query, and therefore these queries were discarded from the final dataset.
We refer to these queries as ``invalid'' for convenience.

We note that just because no relevant passage was found in the top-10 candidates presented to the annotator, it is not necessarily the case that no relevant passage exists in the corpus.
However, we did spot check a few of these ``invalid'' queries using a variety of tools, including interactive searching on the web.
At least based on our limited efforts, we were not able to find relevant passages.
While retrieval models that are able to self-assign confidence to its output (including the special case where the model may believe that no answer exists in the corpus), assessing such a capability is challenging in a retrieval-based setting.
Short of exhaustively evaluating the corpus (obviously impractical), we cannot conclude with certainty that no relevant passage exists.

\section{Baselines}
\label{sec:baselines}

As neural retrieval models have gained in sophistication in recent years, the ``software stack'' for end-to-end systems has grown more complex.
This has increased the barrier to entry for ``newcomers'' who wish to start working on multilingual retrieval.
We believe that the growth of diversity of languages introduced in \miracl should be accompanied by an increase in the diversity of participants.

The need to provide ``easy to use'' starting points and the ongoing quest to make research more reproducible are, we believe, the same challenge.
To this end, our research group has devoted substantial effort in build two toolkits to support reproducible research:\ Anserini~\cite{Yang_etal_JDIQ2018} in Java and 
Pyserini~\cite{Lin_etal_SIGIR2021_Pyserini} in Python.

For \miracl, we make available in Pyserini three baselines to serve as foundations that others can build on.
Baselines scores for these retrieval models are shown in Table~\ref{tab:baselines} in terms of two standard retrieval metrics, nDCG@10 and Recall@100.
In more detail:

\begin{itemize}[leftmargin=*]

\item \textbf{BM25}
\citep{bm25}, one of the retrieval models used in the ensemble system to generate candidate passages for annotation.
\citet{thakur2021beir} and \citet{mrtydi} both show BM25 to be a robust baseline when evaluated zero-shot across domain and languages. 

\item \textbf{mDPR}~\citep{karpukhin-etal-2020-dense, mrtydi}, another one of the retrieval models used in the ensemble system to generate candidate passages for annotation.
DPR is representative of the family of dense retrieval methods that has proven to be effective for many retrieval tasks.

\item \textbf{Hybrid} combines the scores of BM25 and mDPR results.
For each (query, document) pair, the hybrid score is computed as $s_{\textrm{Hybrid}} = \alpha \cdot s_{\textrm{BM25}} + (1 - \alpha) \cdot s_{\textrm{mDPR}}$,
where we set $\alpha=0.5$ without tuning.
Scores of BM25 and mDPR ($s_{\textrm{BM25}}$ and $s_{\textrm{mDPR}}$) are first normalized to $[0, 1]$. 

\end{itemize}

\noindent Code, documentation, and instructions for reproducing these baselines are available at the MIRACL repository, \url{https://github.com/project-miracl/miracl}.
It is our intention to release more baselines as the evaluation progresses.

\section{Ongoing Work}

The broader \miracl project has only begun.
The first step on our journey is the release of the training and development sets in the 16 known languages of \miracl.
These resources are now publicly available on the \miracl website at \url{http://miracl.ai/}.

The next phase of our efforts will focus on finalizing the details of the WSDM 2023 Cup challenge, which will provide a common evaluation methodology, a leaderboard, and a venue for a competition-style event with prizes.
It is our plan to support two tasks: retrieval on the known languages as well as the surprise languages.
In the first case, data have already been released, while in the second case, participants will have only a very limited amount of time to rapidly develop retrieval capabilities.
We are finalizing the ``rules of the game'' as well as the timing of the test set releases.

We invite the community to come join us and help make more MIRACLs!

\section*{Acknowledgments}

This research was supported in part by the Natural Sciences and Engineering Research Council (NSERC) of Canada and a gift from Huawei.
Computational resources were provided in part by Compute Ontario and Compute Canada.

\balance

\bibliography{acmart}
\bibliographystyle{ACM-Reference-Format}

\end{document}